\documentclass[12pt,preprint]{aastex}
\def\kms{km s$^{-1}$}

\def\and{$\&$ }

\begin{document}

\title{Morphologies of Ultracompact HII Regions in W49A and Sgr B2: \\
Prevalence of Shells and a Modified Classification Scheme}

\author{C. G. De Pree\altaffilmark{1}, 
        D. J. Wilner\altaffilmark{2}, 
        J. Deblasio\altaffilmark{1},
	A. J. Mercer\altaffilmark{1,3}, 
        L. E. Davis\altaffilmark{1,4}}

\altaffiltext{1}{Department of Physics and Astronomy, Agnes Scott College, 
                 141 E. College Ave., Decatur, GA 30030, cdepree@agnesscott.edu, jdeblasio@agnesscott.edu}
\altaffiltext{2}{Harvard-Smithsonian Center for Astrophysics, 60 Garden St.,
                 Cambridge, MA 02138, dwilner@cfa.harvard.edu}
\altaffiltext{3}{Department of Physics \& Astronomy, University of Iowa, allison-mercer@uiowa.edu}
\altaffiltext{4}{Department of Astronomy, University of Florida, ledavis@astro.ufl.edu}

\begin{abstract}
We have used Very Large Array (VLA) observations of the massive star forming 
regions W49A and Sgr B2, obtained with resolutions from 2\farcs0 to 0\farcs04,
to classify the morphologies of nearly 100 ultracompact HII regions. 
These high resolution, multi-frequency, multi-configuration VLA observations 
motivate several modifications of the existing morphological classification 
scheme for UC~HII regions. In this work, we describe the modified morphology 
scheme and the criteria used in source classification.  In particular, we 
drop the ``core-halo'' classification, add a ``bipolar'' classification, 
and change the shell classification to ``shell-like''. 
We tally the percentage of each morphology found in the Sgr B2 and W49A 
regions and find broad agreement with the Galactic plane surveys in 
the distribution of morphologies for most types.  However, we find that
nearly a third of the sources in these regions are shell-like, which is a 
higher percentage by nearly a factor of ten than found in the surveys of 
Galactic plane star forming regions by Wood \& Churchwell (1989a) and 
Kurtz et al. (1994). This difference may be due to physical differences in
the environments of these two extreme star forming regions. Alternatively,
differences in observational technique may be responsible.
\end{abstract}

\keywords{interferometry~$-$~nebulae: H~II regions~$-$~nebulae: individual (W49N, Sgr B2)}

\section{Introduction}
Wood \& Churchwell (1989a; hereafter WC89) published the first survey of a 
new class of radio continuum sources in the Galaxy called ultracompact (UC) 
HII regions.  These sources represent a subset of ``standard'' HII regions 
characterized by unusually small sizes ($<$0.01~pc) and high emission measures 
($>$10$^8$~pc~cm$^{-6}$), as derived from high angular resolution imaging 
observations with the Very Large Array (VLA) at 6 and 2 centimeters.
WC89 systematically tabulated the physical properties of 75 sources found
in regions at a variety of distances and placed them in five distinct 
morphological classes: cometary, shell, irregular, core-halo, 
and spherical/unresolved. The percentages of UC~HII region morphologies found 
in this survey were 4\% shell, 20\% cometary, and the remainder (76\%)
irregular, spherical/unresolved and core-halo sources. 
A subsequent survey of 59 additional regions by Kurtz et al. (1994; hereafter 
K94) extended these results and detected 75 more UC~HII regions with a similar 
distribution of morphologies.

Wood \& Churchwell (1989b) estimate the lifetime of an UC~HII region ($\sim$10$^5$ yr) from the comparison between the number of O stars within 2.5 kpc of the sun (436; Conti et al. 1983) and the number of the 1646 embedded OB star candidates from the IRAS sample that one would expect to find within 2.5 kpc of the sun if the sources were uniformly distributed in the Galactic disk (45). From this comparison, they determine that 10\%-20\% 
of the lifetime of an O star is spent in the UC~HII region phase, in agreement with the estimate by Mezger \& Smith (1975) that between 15\% and 25\% of all O stars are hidden by dust. 
Following the arguments of Wood \& Churchwell (1989b), if there are approximately 1700 embedded OB star candidates in the galactic disk, the surveys of Wood \& Churchwell (WC89), Kurtz et al (1994) and the current work sample approximately 20\% of the embedded OB stars in the Galactic disk.

In recent years, we have used the VLA to study radio continuum 
and radio recombination line emission from two of the Galaxy's most 
luminous regions of massive star formation, Sgr B2 and W49A. Other luminous massive star forming regions include the Arches, the Galactic Center cluster, and NGC 3603.
The observations of these ``target rich'' regions, with resolutions 
from $\sim$2'' to 0\farcs04 ($\sim$0.1 to 0.002~pc), have identified
nearly 100 additional UC~HII regions. The imaging observations of this 
new sample motivate a modification to the UC~HII region morphological 
classification scheme of WC89. In particular, we suggest removal of 
the ``core-halo'' morphology, and we add a new ``bipolar'' designation.
In this {\em Letter}, we describe these modifications to the 
classification scheme, the criteria used in the classification process, 
and we discuss the UC~HII region morphologies found in Sgr B2 and W49A.

\section{Observations}
Our VLA radio continuum and recombination line observations of Sgr B2 and 
W49A have been presented and analyzed in a series of previous papers:
7~mm and 1.3~cm observations of Sgr B2 are discussed in 
De Pree et al. (1996, 1998), and Gaume et al. (1995);
7~mm, 1.3~cm, and 3.6~cm observations of W49A are discussed in 
De Pree et al. (1997, 2000) and Wilner et al. (2001);
further details of the high resolution line and continuum imaging of Sgr B2 and W49A are given in De Pree et al. (2004) and De Pree et al. (2005). 

The VLA observations of these two massive star forming regions have provided 
a large sample of UC~HII regions suitable for morphological study.  
The sample has the following characteristics: 
(1) a total number of sources of 97, comparable to the WC89 and K94 surveys, 
(2) sensitivity to radio continuum structures on a wide range of size scales,
(3) good emission measure sensitivity, resulting from long integration times
on a large number of sources within just two fields of view, and
(4) unambiguous relative size measurements within each of the two regions,
as the sources within each region are associated and at the same distance.

\section{Results and Discussion}

\subsection{A Revised Morphological Scheme}
We classified the morphologies of each of the UC~HII regions detected in 
W49A and Sgr B2 using the highest angular resolution images available.
Most of the sources can be placed in one of the morphological classes 
defined by WC89, but characteristics of the data prompts us to modify
the WC89 classification scheme in three ways:

{\em A new bipolar morphology:}
There is a significant population of sources that exhibit a regular, 
previously uncategorized, morphology, elongated along one axis. 
Churchwell (2002), in his recent review of UC~HII regions, refers to a 
``bipolar'' morphology,  and we adopt that term here.  The addition of 
the bipolar morphology is clearly needed to accommodate the presence of
this class of sources. 

{\em Elimination of core-halo morphology:}
With multi-configuration VLA observations that are sensitive to emission
at large size scales, it is apparent that many of the UC~HII regions in 
Sgr B2 and W49A are associated with a faint shelf of radio emission (e.g.
see 3.6 cm images of W49A; De Pree et al. 1997). Recent studies of other,
more isolated, radio continuum sources in star forming regions indicate 
that essentially all UC~HII regions are associated with large scale, 
diffuse emission, when observed with sufficient sensitivity (Kim \& Koo 2003). 
Consequently, the ``core-halo'' designation is not a useful morphological
discriminant in modern data sets of high quality. Instead, we prefer to 
concentrate on the morphology of the compact source within the extended 
emission.

{\em Rename ``shell'' morphology to ``shell-like'':}
Many of the shell-like sources show considerable clumpiness and departures
from circular symmetry in their brightness, while retaining a clearly 
identifiable and approximately circular edge-brightened structure. 
A slight revision in nomenclature to shell-like (SL) offers a better description of this morphology. 

In summary, the revised UC~HII region morphological types are: 
(1) Shell-like (SL), (2) Bipolar, (3) Cometary, 
(4) Spherical, 
(5) Irregular, 
and (6) Unresolved.
The definitions of these morphologies (abbreviations in parenthesis) are:

\begin{itemize}
\item Shell-like (SL): 
Roughly circular sources that are not centrally peaked and are more than 50\% 
complete around their perimeter, with no evidence of elongation away from the 
brightened edge. Figure 1a shows the SL source W49A/D. 

\item Bipolar (B): 
Sources with an axial ratio of at least 2:1. 
If radio recombination line data are available, then there may be evidence of a 
velocity gradient along the long axis. 
Figure 1b shows the bipolar source W49A/A1.

\item Cometary (C): 
Sources with a bright edge and a tail that extends at least 1.5 times the 
width of the face. Figure 1c shows the cometary source Sgr B2 Main/I.

\item Spherical (S): 
Sources that are symmetric, centrally brightened, resolved, and can be 
fit well by a two-dimensional Gaussian. Figure 1d shows the spherical source W49A/R.

\item Irregular (I): 
Any spatially resolved source that does not fit in one of the specified 
categories.  Figure 1e shows the irregular source Sgr B2 Main/F2

\item Unresolved (U): Sources that are symmetrical, centrally brightened, 
and can be fit well by a two-dimensional Gaussian that is comparable to 
the beam size. Figure 1f shows the unresolved source W49A/A2

\end{itemize}

\subsection{Bipolar Sources}
The newly recognized class of bipolar UC~HII regions is especially interesting.
Kinematic evidence from radio recombination lines suggest that the bipolar 
morphology may be associated with the bipolar outflow phase of early stellar 
evolution.  Several of the bipolar sources in W49A and Sgr B2 show velocity 
gradients along their long axis, or they show remarkably broad line widths 
($\Delta$V$_{FWHM}>$50 \kms) that are not the result of pressure broadening (De Pree et al. 2004). The motions of the ionized gas, together with 
the elongation along one axis, suggest that ionized gas is flowing away from 
the young massive star at highly supersonic velocities. 
In addition, Garay et al. (2004) have observed a collimated stellar wind emanating from 
IRAS 16547-4247 and conclude that the thermal jet phenomenon may be common for 
high-mass as well as for low-mass stars.

\subsection{Prevalence of Shells}
Table 1 lists the percentage of sources found in each morphological class 
for the sources in the surveys of WC89, K94, and in SgrB2 and W49.
As indicated in Table 1, we find that 28\% are shell-like, 8\% are bipolar, 
14\% are cometary, and the remainder(50\%) are spherical, irregular, 
or unresolved. 
In most respects, the findings of the three surveys are compatible within
the uncertainties, which are generally dominated by counting statistics
(and the subjective nature of the classification process). The frequencies 
of cometary, irregular and unresolved morphologies are compatible.
K94 find a marginally higher percentage of spherical sources than in the 
WC89 survey or the W49A and Sgr B2 sample. The bipolar morphology was not 
used by WC89 or K94, and we do not use the core-halo morphology, thus we
do not have percentages to compare for these two morphologies.

The only significant difference between the morphological classifications 
within Sgr B2 and W49A and those of WC89 and K94 is the much higher 
percentage of SL sources (28\% vs. 3\%), about an order of magnitude more
frequent occurrence of sources in this class. 
The rarity of shell-like sources in the earlier surveys is surprising, considering the results of existing hydrodynamical models. Studies of the evolution of UC~HII regions that examine the expansion of photoionization and stellar wind bubbles in dense molecular gas suggest that shell-like morphologies should be common (e.g. Garcia-Segura \& Franco 1996). The VLA observations of the Sgr B2 and W49A regions indicate that SL is the most commonly observed morphology in these regions, comprising approximately one-third of the sources.

There are several possible explanations for the striking difference in the 
percentage of shell sources found by WC89, K94, and in the current work. Since 
the difference has clear statistical significance, it must result either 
from real physical differences in the populations of UC~HII regions sampled, 
or from selection effects related to differences in the observational strategies of the Galactic survey work (WC89 and K94) and the current work.

The star forming environments represented by Sgr B2 and W49A are extreme
and present differences from the typical Galactic plane star forming 
environment. Since the WC and K94 surveys targeted isolated sources identified 
from {\em IRAS} 100 $\micron$ measurements, these surveys contained primarily 
isolated star forming regions, rather than highly clustered, high source 
density regions like Sgr B2 and W49A. 
The difference in the percentage of shell-like sources could indicate that 
physical conditions in these highly clustered regions lead to a higher 
prevalence of this morphology. 
One possibile explanation for the larger numbers of regular shell-like morphologies is
that the molecular environments into which the UC HII regions expand
are smoother, with fewer fluctuations in ambient pressure due to density,
temperature, or turbulence that lead to the development of substantial
asymmetries in the expansion phase.

The observing strategies and goals of WC89 and K94 differed
from the current work.  The WC89 and K94 surveys were comprised of
short ``snapshot'' observations that targeted a large number of relatively
isolated regions at the expense of integration time on each region. 
The Sgr B2 and W49A surveys, by contrast, consist of observations of just two 
(crowded) pointings, which allowed for substantially longer integration times, 
better Fourier plane coverage, and higher sensitivity. Given data with
lower sensitivity, many of the 
sources identified in Sgr B2 and W49A with SL morphology might have been 
classified as cometary or irregular. 
Likewise, sources classified by WC89 and K94 as 
cometary and irregular might (with better sensitivity and multiple VLA configurations) be classified as shell-like. 
Fey et al. (1992) show several examples of such changes in UC~HII region
morphological classification in light of data of higher quality and 
sensitivity to a broader range of spatial scales. 

\section{Conclusions}
We have classified the morphologies of nearly 100 UC~HII regions in the Sgr B2 
and W49A regions. These results, taken together with previous work, suggest
modifications in the UC~HII region classification scheme, as follows:
\begin{enumerate}

\item VLA observations with sensitivity to a wide range of spatial scales 
detect faint extended radio continuum emission throughout massive star forming 
regions, which makes nearly all compact sources appear with ``core-halo'' 
morphologies. As a result, we drop the ``core-halo'' designation from the
UC~HII region classification scheme.

\item A small percentage of UC~HII regions have clear bipolar morphologies; in 
cases where recombination line data are available, the kinematics suggest 
directed ionized outflows originating from a young central source. This new 
``bipolar'' morphological class is more common in W49A and Sgr B2 than the 
shell morphology was in the WC89 or K94 surveys.

\item We suggest that the shell classification should be modified to 
``shell-like'' (or SL), in order to describe more accurately 
the appearance of these sources, many of which are clumpy or partially complete.
The Sgr B2 and W49A regions exhibit $\sim$10 times as many sources 
with shell-like morphologies as found in the WC89 or K94 surveys. Deep,
multi-configuration VLA observations of sources previously classified as
cometary and irregular should determine whether this discrepancy is due 
to limitations of snapshot data.  
\end{enumerate}

The next step in this study is to utilize the morphological findings and 
derived physical parameters of the sources and their environments as inputs
and constraints in hydrodynamic models of UC~HII region evolution.

\acknowledgements{
This research is supported by a grant from the National Science Foundation 
(AST-0206103). The authors gratefully acknowledge E. Churchwell and M. Goss for making helpful comments on an early draft of this paper. CGD thanks Emory University Department of Physics for its hospitality during his sabbatical in Fall 2004.}

\clearpage

\begin{figure}[p]
\vspace{18cm}
\includegraphics{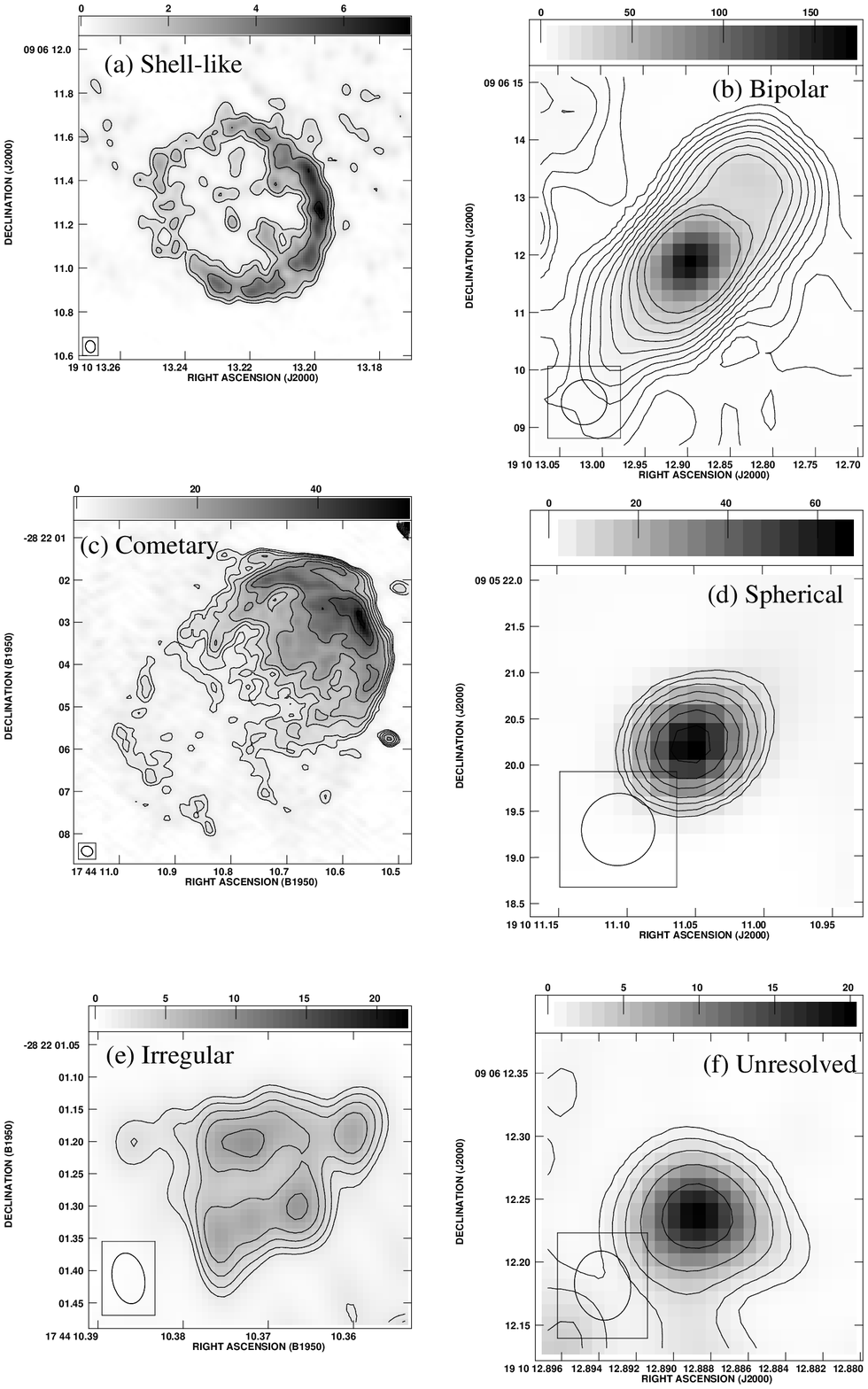}
\caption{Sample morphologies, with observing frequency and original reference in parenthesis (a) Shell-like (SL) source, W49A/D (7 mm, De Pree et al. 2004), (b) Bipolar (B) source, W49A/A (3.6 cm, De Pree et al. 1997) (c) Cometary (C) source, Sgr B2 Main/I (1.3 cm, Gaume et al. 1995) (d) Irregular (I) source, Sgr B2 Main/F2 (7 mm, De Pree et al. 1998), (e) Spherical (S) source, W49A/R (3.6 cm, De Pree et al. 1997), (f) Unresolved (U) source, W49A/A2 (7 mm, De Pree et al. 2004)}
\end{figure}

\clearpage

\begin{table}
\begin{center}
\begin{tabular}{lccc}   \hline \hline
Morphology & SgrB2/W49 & WC (1989) & Kurtz et al.(1994) \\ \hline
Shell-like (SL) & 28 & 4\tablenotemark{a} & 1\tablenotemark{a}\\
Bipolar (B) & 8 & \nodata\tablenotemark{b} & \nodata\tablenotemark{b}\\
Cometary (C) & 14 & 20 & 16\\
Core-Halo\tablenotemark{c} (CH) & \nodata\tablenotemark{c} & 16 & 9\\
Irregular (I) & 11 & 17 & 19\\
Spherical (S) & 21 & 24 & 36\\
Unresolved (U) & 18 & 19 & 19\\
\hline\hline
\end{tabular}
\tablenotetext{a}{WC89 and K94 called this simply a shell morphology.}
\tablenotetext{b}{WC89 and K94 did not include a bipolar morphology.}
\tablenotetext{c}{WC89 and K94 included a core-halo morphology (16\% of their 
survey sources); we eliminate this designation and instead concentrate on the 
morphology of the compact core (see text).}
\end{center}
\tablenum{1}
\caption{Morphology Distributions}
\end{table}






\end{document}